%% file: prochiggs.tex
\documentclass[12pt]{article}
\usepackage{graphicx}
\usepackage{cite}

\def\Kenensaw{Department of Physics\\
Kennesaw State University, Kennesaw, GA 30144, USA }
\def\support{\footnote{This material is based upon work supported by the National Science Foundation under Grant No. PHY 1519606.}}

\def\Title#1{\begin{center} {\Large #1 } \end{center}}
\def\Author#1{\begin{center}{ \sc #1} \end{center}}
\def\Address#1{\begin{center}{ \it #1} \end{center}}

\newenvironment{Abstract}{\begin{quotation}  }{\end{quotation}}
\newenvironment{Presented}{\begin{quotation} \begin{center} 
             PRESENTED AT\end{center}\bigskip 
      \begin{center}\begin{large}}{\end{large}\end{center} \end{quotation}}

\input econfmacros.tex
\def\beq{\begin{equation}}
\def\eeq{\end{equation}}
\def\beqa{\begin{eqnarray}}
\def\eeqa{\end{eqnarray}}

\begin{document}
\begin{titlepage}

\vfill
\Title{Theoretical results for charged-Higgs production}
\vfill
\Author{Nikolaos Kidonakis\support}
\Address{\Kenensaw}
\vfill
\begin{Abstract}
I discuss charged-Higgs production via two different processes: in association with a top quark, and in association with a $W$ boson. I present total cross sections and differential distributions that include higher-order corrections from soft and collinear gluon emission through aN$^3$LO. I show that these radiative corrections are significant.
\end{Abstract}
\vfill
\begin{Presented}
CIPANP2018\\
Palm Springs, California, May 29--June 3, 2018
\end{Presented}
\vfill
\end{titlepage}
\def\thefootnote{\fnsymbol{footnote}}
\setcounter{footnote}{0}

\section{Introduction}

Since the discovery of the (neutral) Higgs boson a lot of attention has been 
given to its properties and the determination whether it is the Standard Model Higgs.
However, any discovery of a charged Higgs boson would be evidence of new physics, 
and the LHC can observe or exclude such a possibility in a wide mass range. 

Here I present results for two distinct production processes of charged Higgs bosons  
in the MSSM (or other 2-Higgs doublet models).
I will discuss the processes
$bg \rightarrow t H^-$  and $b{\bar b} \rightarrow H^- W^+$.

Higher-order QCD corrections are significant for both processes. Furthermore, 
since the processes involve very massive final states, soft-gluon corrections are important 
and constitute the bulk of the corrections. Below I present theoretical results for these
corrections, and numerical results for cross sections and distributions at LHC energies.

\section{$tH^-$ production}

\begin{figure}[htb]
\centering
\includegraphics[height=1.5in]{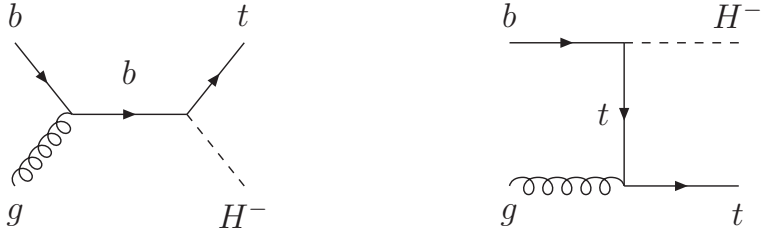}
\caption{Lowest-order diagrams for $bg \rightarrow tH^-$.}
\label{tHplot}
\end{figure}

The leading-order cross section for $bg \rightarrow t H^-$ is proportional to  
$\alpha \alpha_s (m_b^2\tan^2 \beta+m_t^2 \cot^2 \beta)$, 
where $\tan \beta=v_2/v_1$ is the ratio of the vacuum expectation values of the two Higgs doublets. The lowest-order diagrams are shown in Fig. \ref{tHplot}.

Fixed-order QCD and SUSY-QCD corrections through NLO have been calculated 
for this process in Refs. \cite{BGGS,SHZ,GLXY,TP,BHJP,WMZJHG,DKSW,FKKSU,DUWZ}. 
Soft-gluon terms constitute a numerically dominant piece of the QCD corrections. 

With the momenta assignments, 
$b(p_b) + g(p_g) \longrightarrow t(p_t)+H^-(p_H)$, 
we define the usual variables $s=(p_b+p_g)^2$, $t=(p_b-p_t)^2$, $u=(p_g-p_t)^2$, 
and furthermore the threshold variable $s_4=s+t+u-m_t^2-m_H^2$ which goes to 0 
at partonic threshold. Soft-gluon corrections appear in the cross section as logarithms of the form $[\ln^k(s_4/m_H^2)/s_4]_+$ \cite{NKtH,NKtHW,NKtH16}.

We resum these soft corrections for the double-differential cross section
at NNLL accuracy, using two-loop soft anomalous dimensions 
\cite{NKtHW,NKtH16,NKloop}.
Taking moments of the partonic cross section,
${\hat \sigma}(N)=\int (ds_4/s) \; e^{-N s_4/s} {\hat \sigma}(s_4)$,  
we write the factorized expression for the dimensionally regularized 
cross section 
\beq
{\hat \sigma}^{bg \rightarrow tH^-}(N,\epsilon)= 
\left( \prod_{i=b,g} J_i\left (N,\mu,\epsilon \right) \right)
H^{bg \rightarrow tH^-} \left(\alpha_s(\mu)\right)\; 
S^{bg \rightarrow tH^-} \left(\frac{m_H}{N \mu},\alpha_s(\mu) \right)
\eeq
where $J_i$ denote functions for the incoming $b$-quark and gluon, 
$H^{bg\rightarrow tH^-}$ is the hard function, 
and $S^{bg\rightarrow tH^-}$ is the soft function.

The soft anomalous dimension $\Gamma_S^{bg \rightarrow tH^-}$ controls the evolution
of $S^{bg\rightarrow tH^-}$, resulting in the exponentiation of logarithms of $N$.
Writing the perturbative series for $\Gamma_S^{bg \rightarrow tH^-}$ as
\beq
\Gamma_S^{bg \rightarrow tH^-}=\frac{\alpha_s}{\pi} \Gamma_S^{(1)}+\left(\frac{\alpha_s}{\pi}\right)^2 \Gamma_S^{(2)}+\cdots \, , 
\eeq 
a one-loop calculation gives \cite{NKtH}
\beq
\Gamma_S^{(1)}=C_F \left[\ln\left(\frac{m_t^2-t}{m_t\sqrt{s}}\right)
-\frac{1}{2}\right] +\frac{C_A}{2} \ln\left(\frac{m_t^2-u}{m_t^2-t}\right)
\eeq
while a two-loop calculation gives \cite{NKtHW}
\beq
\Gamma_S^{(2)}=\left[C_A \left(\frac{67}{36}-\frac{\zeta_2}{2}\right)-\frac{5}{18} n_f\right] \Gamma_S^{(1)} +C_F C_A \frac{(1-\zeta_3)}{4} \, .
\eeq

We then expand the resummed cross section and invert to momentum space, thus deriving approximate cross sections at NNLO and N$^3$LO. 
The approximate NNLO (aNNLO) soft-gluon corrections are 
\beq
\frac{d^2{\hat \sigma}_{\rm aNNLO}^{(2) \, bg \rightarrow tH^-}}{dt \, du}=F_{\rm LO}^{bg \rightarrow tH^-} \frac{\alpha_s^2}{\pi^2} 
\sum_{k=0}^3 C_k^{(2)} \left[\frac{\ln^k(s_4/m_H^2)}{s_4}\right]_+
\eeq
with coefficients
$C_3^{(2)}=2(C_F+C_A)^2$ and 
\beqa
C_2^{(2)}&=&(C_F+C_A) \left\{3C_F\left[2\ln\left(\frac{m_t^2-t}{m_t \sqrt{s}}\right)-2\ln\left(\frac{m_H^2-u}{m_H^2}\right)-1\right] \right.
\nonumber \\ && \left. \hspace{-8mm} 
{}-3 C_A \left[\ln\left(\frac{m_t^2-t}{m_t^2-u}\right)
+2 \ln\left(\frac{m_H^2-t}{m_H^2}\right)\right]
-3(C_F+C_A) \ln\left(\frac{\mu_F^2}{s}\right)-\frac{\beta_0}{2} \right\} \, .
\eeqa
The expressions for $C_1^{(2)}$ and $C_0^{(2)}$ are much longer \cite{NKtHW}.

The approximate N$^3$LO (aN$^3$LO) soft-gluon corrections are:
\beqa
\frac{d^2{\hat \sigma}_{\rm aN^3LO}^{(3) \, bg \rightarrow tH^-}}{dt \, du}
=F_{\rm LO}^{bg \rightarrow tH^-} \frac{\alpha_s^3}{\pi^3} 
\sum_{k=0}^5 C_k^{(3)} \left[\frac{\ln^k(s_4/m_H^2)}{s_4}\right]_+
\eeqa
with coefficients
$C_3^{(3)}=\frac{1}{2}(C_F+C_A)^3$, etc.

We now present numerical results for $tH^-$ production at LHC energies. We use MMHT2014 NNLO pdf \cite{MMHT2014} in our calculations.

\begin{figure}[htb]
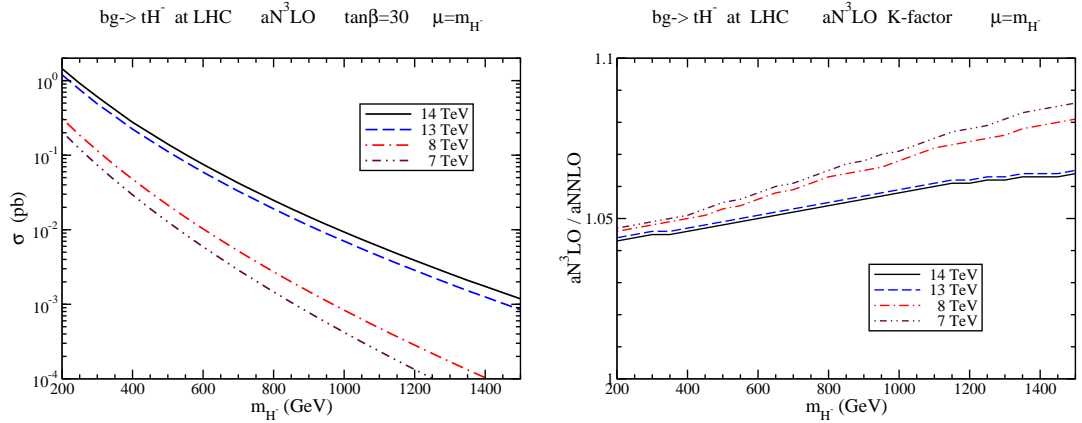

\centering
\includegraphics[height=2.2in]{chiggsaN3LOtn30plot.eps}
\hspace{2mm}
\includegraphics[height=2.2in]{KaN3LOchiggslhcplot.eps}
\caption{Total aN$^3$LO cross sections (left) and aN$^3$LO /aNNLO ratios (right) for $tH^-$ production.}
\label{chiggsaN3LO}
\end{figure}
The aN$^3$LO cross sections at LHC energies are plotted in Fig. \ref{chiggsaN3LO}. The left plot gives the total cross sections at each LHC energy as functions of charged Higgs mass while the plot on the right shows the aN$^3$LO/aNNLO ratios.   

\begin{figure}[htb]
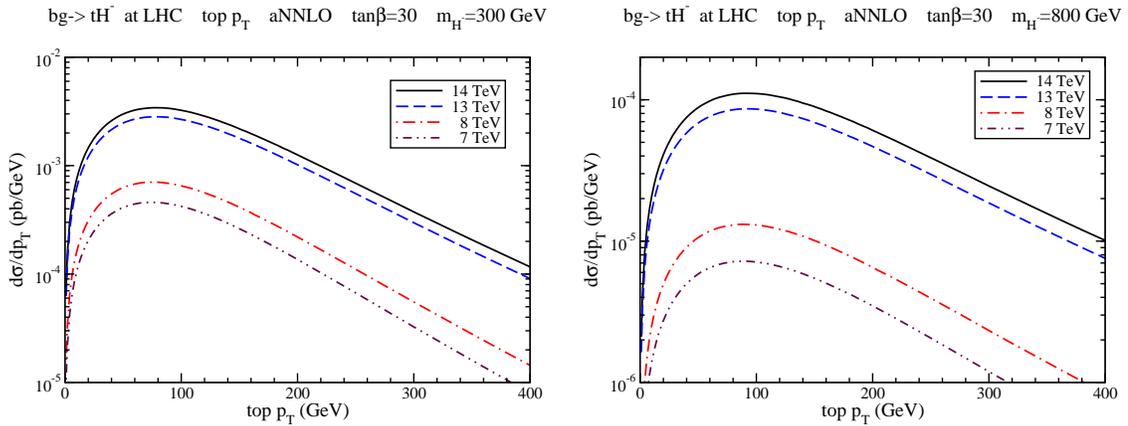

\centering
\includegraphics[height=2.2in]{pttopchiggs300tn30plot.eps}
\hspace{2mm}
\includegraphics[height=2.2in]{pttopchiggs800tn30plot.eps}
\caption{Top-quark $p_T$ distributions at aNNLO in $tH^-$ production for a charged Higgs mass of (left) 300 GeV and (right) 800 GeV.}
\label{ptchiggsaNNLO}
\end{figure}
The top-quark aNNLO transverse-momentum, $p_T$, distributions at LHC energies are plotted in Fig. \ref{ptchiggsaNNLO}. The left plot is for a charged Higgs mass of 300 GeV, while the right plot uses 800 GeV.

\begin{figure}[htb]
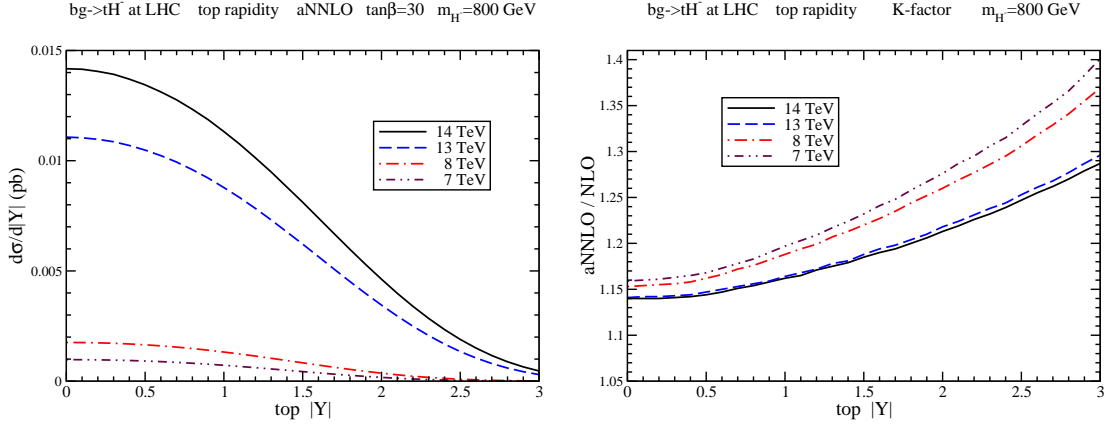

\centering
\includegraphics[height=2.2in]{yabstopchiggs800tn30plot.eps}
\hspace{2mm}
\includegraphics[height=2.2in]{Kyabstopchiggs800plot.eps}
\caption{Top-quark rapidity distributions at aNNLO (left) and aNNLO/NLO ratios (right) in $tH^-$ production for a charged Higgs mass of 800 GeV.}
\label{ychiggsaNNLO}
\end{figure}
The top-quark aNNLO rapidity distributions at LHC energies are plotted in Fig.~\ref{ychiggsaNNLO} for a charged Higgs mass of 800 GeV. As the aNNLO/NLO ratios in the plot on the right show, the higher-order soft-gluon corrections are large, and increase sharply for larger values of rapidity.

The fact that soft-gluon corrections are large and dominate the corrections for $tH^-$ production is also consistent with analogous results for $t{\bar t}$ production \cite{N3LOtt,NKtop} and single-top \cite{NKtHW,NKtop,NKsingletop} production.

\section{$H^-W^+$ production}

\begin{figure}[htb]
\centering
\includegraphics[height=1.5in]{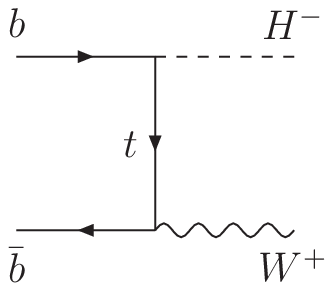}
\caption{Lowest-order diagram for $b{\bar b} \rightarrow H^- W^+$.}
\label{HWdiag}
\end{figure}

We continue with $H^- W^+$ production via the process  
\beq
b(p_1)\, + \, {\bar b}\, (p_2) \rightarrow H^-(p_3)\, + W^+(p_4) \, ,
\eeq
for which the lowest-order diagram is shown in Fig. \ref{HWdiag}.

Fixed-order radiative corrections were calculated for this process in 
Refs. \cite{DHKR,DK,YLJZ,ZMJHW,BHK,HZ,ABK,ZLL,EHR,GLL,MH,BTW,NHN,AGKMSY,ARD,BGLLS,EPS,LWY}. Soft-gluon contributions are a numerically large and dominant part of higher-order corrections.

Again, defining $s=(p_1+p_2)^2$, $t=(p_1-p_3)^2$, $u=(p_2-p_3)^2$ 
and $s_4=s+t+u-m_H^2-m_W^2$, the soft-gluon corrections appear 
as terms of the form $[\ln^k(s_4/m_H^2)/s_4]_+$ \cite{NKHW}.
In addition, we include collinear terms of the form $(1/m_H^2) \ln^k(s_4/m_H^2)$
since they are numerically important, as was also found for the related process
$b{\bar b} \rightarrow H$ in Ref. \cite{NKbbH}.

We write again a factorized expression for the cross section as
\beq
{\hat \sigma}^{b{\bar b} \rightarrow H^- W^+}(N,\epsilon)= 
\left( \prod_{i=b,{\bar b}} J_i\left (N,\mu,\epsilon \right) \right)
H^{b{\bar b} \rightarrow H^- W^+} \left(\alpha_s(\mu)\right)\; 
S^{b{\bar b} \rightarrow H^- W^+} \left(\frac{m_H}{N \mu},\alpha_s(\mu) \right) \, ,
\eeq
where $J_i$ denote functions for the incoming $b$ and ${\bar b}$ quarks, 
$H^{b{\bar b} \rightarrow H^- W^+}$ is the hard function, 
and $S^{b{\bar b} \rightarrow H^- W^+}$ is the soft function.
We perform the resummation of collinear and soft-gluon corrections, 
and expand to fixed order.

The aNNLO collinear and soft-gluon corrections are \cite{NKHW}
\beq
\frac{d^2{\hat{\sigma}}_{\rm aNNLO}^{(2) \, b{\bar b} \rightarrow H^- W^+}}{dt \, du}= F_{LO}^{b{\bar b} \rightarrow H^- W^+} 
\frac{\alpha_s^2}{\pi^2}
\left\{-C_3^{(2)} \frac{1}{m_H^2} \ln^3\left(\frac{s_4}{m_H^2}\right) 
+\sum_{k=0}^3 C_k^{(2)} \left[\frac{\ln^k(s_4/m_H^2)}{s_4}\right]_+ \right\}
\eeq
with $C_3^{(2)}=8 C_F^2$
\beq
C_2^{(2)}=-12 C_F^2\left(\ln\left(\frac{(t-m_W^2)(u-m_W^2)}{m_H^4}\right)
+\ln\left(\frac{\mu_F^2}{s}\right)\right)  
-\frac{11}{3} C_F C_A +\frac{2}{3} C_F n_f \, .
\eeq
The expressions for $C_1^{(2)}$ and $C_0^{(2)}$ are much longer and are given in 
\cite{NKHW}.

\begin{figure}[htb]
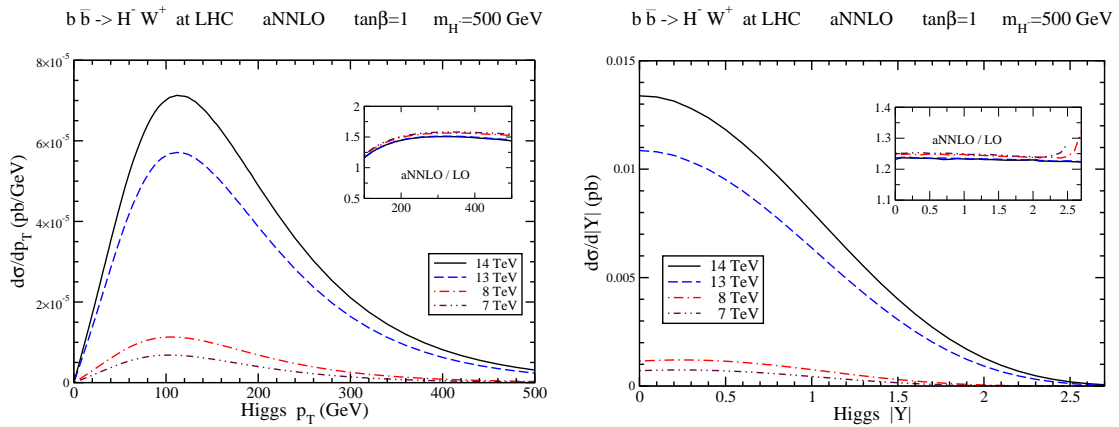

\centering
\includegraphics[height=2.2in]{pthiggsbbHWaNNLO500tan1plot.eps}
\hspace{2mm}
\includegraphics[height=2.2in]{yabshiggsbbHWaNNLO500tan1plot.eps}
\caption{The charged-Higgs aNNLO $p_T$ (left) and rapidity (right) distributions in $H^- W^+$ production.}
\label{pTyHWplot}
\end{figure}

Next, we present results for $H^-W^+$ production at LHC energies. We use the MMHT2014 NNLO pdf \cite{MMHT2014} in our calculations but note that results using CT14 pdf \cite{CT14} are nearly the same.

In Fig. \ref{pTyHWplot} we present the charged-Higgs $p_T$ and rapidity distributions at aNNLO. 
The inset plots show the aNNLO/LO ratios, which show that the soft-gluon corrections are very significant. 
These results are also in line with analogous results for $W$ production that were presented in \cite{WZ}. 

The aN$^3$LO collinear and soft-gluon corrections are
\beq
\frac{d^2{\hat{\sigma}}_{\rm aN^3LO}^{(3) \, b{\bar b} \rightarrow H^- W^+}}{dt \, du}= F_{LO}^{b{\bar b} \rightarrow H^- W^+} 
\frac{\alpha_s^3}{\pi^3}
\left\{-C_5^{(3)} \frac{1}{m_H^2} \ln^5\left(\frac{s_4}{m_H^2}\right) 
+\sum_{k=0}^5 C_k^{(3)} \left[\frac{\ln^k(s_4/m_H^2)}{s_4}\right]_+ \right\}
\eeq
with $C_5^{(3)}=8 C_F^3$, etc. \cite{NKHW}.

The aN$^3$LO corrections are numerically small but they have big uncertainties. Therefore we do not show numerical results for them here.

\section{Summary}

I have presented new results through aN$^3$LO for charged Higgs production in association with a top quark or a $W$ boson.
Soft-gluon corrections have been derived from NNLL resummation at aNNLO and aN$^3$LO.

Total cross sections for $tH^-$ production have been presented at aN$^3$LO and it is shown that 
the soft-gluon corrrections are large.
Top-quark $p_T$ and rapidity distributions in $tH^-$ production were also presented at aNNLO.

I have also presented cross sections and charged-Higgs $p_T$ and rapidity distributions in $H^- W^+$ production at aNNLO.
The higher-order soft-gluon corrections are very significant at LHC energies.

\end{document}

%% file: econfmacros.tex
\def\beq{\begin{equation}}
\def\eeq#1{\label{#1}\end{equation}}
\def\eeqn{\end{equation}}

\def\beqa{\begin{eqnarray}}
\def\eeqa#1{\label{#1}\end{eqnarray}}
\def\eeqan{\end{eqnarray}}

\let\bar=\overbar

\def\Dslash{\not{\hbox{\kern-4pt $D$}}}
\def\dslash{\not{\hbox{\kern-2pt $\del$}}}

\def\msb{{\bar{\ssstyle M \kern -1pt S}}}

